\documentclass[twocolumn,showpacs,preprintnumbers,superscriptaddress,amsmath,amssymb,prl]{revtex4}

\usepackage{graphicx}
\usepackage{dcolumn}
\usepackage{bm}
\usepackage{amsmath}

\newcommand{\be}{\begin{equation}}
\newcommand{\ee}{\end{equation}}
\newcommand{\ba}{\begin{eqnarray}}
\newcommand{\ea}{\end{eqnarray}}

\begin{document}

\preprint{APS preprint}

\title{Self-fulfilling Ising Model of Financial Markets}

\author{W.-X. Zhou}
\affiliation{State Key Laboratory of Chemical Reaction Engineering,\\
East China University of Science and Technology, Shanghai 200237,
China}

\author{D. Sornette}

\email{sornette@moho.ess.ucla.edu}

\affiliation{Institute of Geophysics and Planetary Physics and
Department of Earth and Space Sciences, University of California,
Los Angeles, CA 90095} \affiliation{Laboratoire de Physique de la
Mati\`ere Condens\'ee, CNRS UMR 6622 and Universit\'e de
Nice-Sophia Antipolis, 06108 Nice Cedex 2, France}

\date{\today}

\begin{abstract}

We study a dynamical Ising model of agents' opinions (buy
or sell) with coupling coefficients reassessed continuously in time
according to how past external news (magnetic field) have explained realized market returns.
By combining herding, the impact of external news and
private information, we test within the same model
the hypothesis that agents are rational versus irrational. We find
that the stylized facts of financial markets are
reproduced only when agents are over-confident and mis-attribute the
success of news to predict return to herding effects, thereby providing
positive feedbacks leading to the model functioning close to the
critical point.

\end{abstract}


\maketitle

Social systems offer a fascinating field for the application of recent
concepts and methods developed in Physics to tackle complex $N$-body
systems with nonlinear feedbacks, and many competing states. A long
tradition started with the application of Ising models and its
extensions to social interactions and organization
\cite{PhysicsToday,Montroll,Galam82,Orlean6}.
A large set of economic models can be
mapped to various versions of the Ising model to account for social
influence in individual decisions (see \cite{Phanetal}
and references therein). Other recent works using the Ising
model include models of bubbles and crashes \cite{JLS00IJTAF,Kaizoji}, a
version with stochastic coupling coefficients which leads to volatility
clustering and a power law distribution of returns at a single fixed
time scale \cite{prlrandis}, and models of opinion polarization
\cite{Galamso,Staufferso}. The dynamical
updating rules of the Ising model can be shown to describe the formation
of decisions of boundedly rational agents \citep{RS00} or to result from
optimizing agents whose utilities incorporate a
social component \cite{Phanetal}. The Ising  model is one of the
simplest models describing the competition between the ordering force of
imitation or contagion and the disordering impact of private information
or idiosyncratic noise, which leads already to the crucial
concept of spontaneously symmetry breaking and phase transitions \cite{Ising1}.

However, human beings
are not spins, they can learn, that is, adapt the nature and strength
of their interactions with others, based on past experience. In the language of the
Ising model, this amounts to generalize to time-dependent
coupling coefficients which reflect past experience.
Here, we study a generalized Ising model of interacting agents buying and selling a
single financial asset who base their decisions
on a combination of mutual influences or imitation,
external news and idiosyncratic judgements. The novel
ingredient is that they update their
willingness to extract information from the other agents' behavior
based on their assessment of how past news have explained market
returns. Agents update their propensity to herding according to
how the news have been successful in predicting returns.
We  distinguish between two possible updating rules: rational and
irrational. In the rational version, agents decrease their
propensity to imitate if news have been good predictors of returns
in the recent past. In the irrational version, agents
mis-attribute the recent predictive power of news to their
collective action, leading to positive self-reinforcement of
imitation. We show that the model can reproduce the major
empirical stylized facts of financial stock markets only when the
updating of the strength of imitation is irrational, providing a
direct test and the evidence for the importance of misjudgement of
agents biased toward herding. This model also offers a dynamical
derivation of the multifractal properties of the structure functions of
the absolute values of returns and their consequences in the
characteristic power law relaxations of the volatility after bursts of
endogenous versus exogenous origins.

Consider $n^2$ agents interacting within a $n\times n$ 2D
square lattice network $\mathcal{N}$ (we have verified
that the properties described below are not sensitive to $n$
in the range $20-100$ that we tested). At each time step $t$, agent
$i$ places a buy ($s_i(t)=+1$) or sell ($s_i(t)=-1$) order
according to the following process
\begin{equation}
 s_i(t)={\rm{sign}}\left[\sum_{j\in{\mathcal{N}}}K_{ij}(t) {\rm E}[s_j](t)+
 \sigma_i(t) G(t) +\epsilon_i(t)\right]~,
 \label{Eq:Sit22}
\end{equation}
where ${\rm E}[s_j](t)$ is the expectation formed by agent $i$ on
what will be the decision of agent $j$ at the same time $t$. An
agent $i$ imitates only her friends, that is $K_{ij}=0$ if $i=j$
or $j$ is not a friend of (connected to) $i$. Expression
(\ref{Eq:Sit22}) embodies three contributions: (i) Imitation in
which $K_{ij}$ is the relative propensity of the trader $i$ to be
contaminated by the sentiment of her friend $j$; (ii) The impact
$G(t)$ of external news (positive resp. negative for favorable
resp. unfavorable news) and $\sigma_i(t)$ is the relative
sensitivity of agent's sentiment to the news, uniformly
distributed in the interval $(0,\sigma_{\max})$ and frozen to
represent the heterogeneity of the agents; (iii) Idiosyncratic
judgement $\epsilon_i(t)$ associated with private information,
assumed to be normally distributed around zero with an
agent-dependent standard deviation $s_{\epsilon,i}$ equal to the
sum of a common constant $CV$ and of a uniform random variable in
the interval $[0,0.1]$ again to capture the heterogeneity of
agents. We have tested several implementation of the formation of
expectations ${\rm E}[s_j](t)$ in (\ref{Eq:Sit22}), such as
backward looking (${\rm E}_i[s_j](t) = s_j(t-1)$ for all $i$ and
$j$'s) or information cascades along specific chains within the
network \cite{G03AFM}, which give similar results.

We account for the adaptive nature of agents and their learning
abilities by updating the coefficient of influence of agent $j$ on
agent $i$ according to the following rule:
\begin{equation}
K_{ij}(t) = K_{i}(t) = b_i +  \alpha K_{i}(t-1) + \beta r(t-1)
G(t-1) ~.
 \label{Eq:RevASM:Ki}
\end{equation} The idiosyncratic imitation tendency $b_i$ of agent $i$
is uniformly distributed in $(0,b_{\max})$ and frozen. The coefficient
$\alpha$ quantifies the persistence of past influences on
the present (the case $\alpha=0$ has
been studied in part in Ref.~\cite{G03AFM}). It captures
the fact that social connections
evolve slowly and exhibit significant persistence, as documented
in numerous studies \cite{Suitor}. Networks of investors communicating
their opinions and sentiment on the stock market are similarly
persistent. The last term with $\beta \neq 0$
quantifies how agent $i$ updates her propensity for imitation based on
the role of the exogenous news $G(t)$ in determining the sign and
amplitude of the observed return in the preceding time period.
Note that, by construction, the model
is non-variational as $K_{ji}(t) \neq K_{ij}(t)$ in general
if the $b_i$'s are different, but this
is not crucial for our results.
Finally, the market price is updated according to $p(t)=p(t-1) \exp[r(t)]$ where
the return $r(t)$ obeys $r(t) =
\frac{\sum_{i\in{\cal{N}}}s_i(t)}{\lambda N}$, where $\lambda$
measures the market depth or liquidity. The return is thus proportional
to the ``magnetization'' or aggregated decisions of the agents.

The sign of the coefficient $\beta$ is crucial.
For $\beta <0$, agent $i$ is less and less influenced by
other agents, the better has been the success of the news in
determining the direction and amplitude of the market return. This
process is self-reinforcing since, as $K_{ij}$ decreases, the
dominant term becomes $ \sigma_i(t) G(t)$, which further ensures
that the news correctly predict the decision of agents and
therefore the direction of the market move, thus decreasing
further the coefficient of influence $K_{ij}$. Reciprocally,
agents tends to be more influenced by others when the news seems
to incorrectly predict the direction of the market. The news being
not reliable, the agents turn to other agents, believing that
others may have useful information. This is in agreement with
standard economics which views the
stock market as an efficient machine transforming all news into prices.

For $\beta >0$, the more the news predict the direction of
the market, the more the agents imitate other agents.
This is the ``irrational'' case where agents either
mis-attribute the origin of the market moves to herding rather
than to the impact of news, or misinterpret the exogenous
character of news in terms of endogenous herding or infer that
other agents will be following more eagerly as a group the
direction given by the news. This may occur due to
mutually-reinforcing optimism \cite{heathgonz} and overconfidence
\cite{Darkefreedman}.

In our simulations, we fix $\lambda =40$ to obtain
returns with amplitudes comparable to that of empirical
observations and $\alpha=0.2$. Similar results are
obtained for $\alpha=0.4$, $0.6$, and $0.8$.
We have explored the properties of the model in the parameter space of
$b_{\max}$, $\sigma_{\max}$ and $CV$. There is no loss of
generality in fixing $|\beta|=1$ to explore the relative
importance of the term $\beta r(t-1) G(t-1)$, since the typical
scale of the $K_i$'s is set by $b_{\max}$ whose amplitude is
varied in our numerical exploration.

For $\beta=-1$, it is easy to
show that the attractor of the dynamics is characterized by
negligible imitation and only the news and private information terms are
important for the dynamics. Indeed, starting from large $K_i(t)$'s
such that the system is above its critical value $K_c$
and the agents behave collectively (ferromagnetic phase), the news
$G(t)$ acts as a ``magnetic'' field which orders the agents'
decision accordingly, leading to the news correctly predicting
the returns. Since $\beta=-1$, the coupling coefficients
$K_i$'s will be decreased by the amount $|r(t)|$. This will
continue until the $K_i$'s are in majority much smaller than $K_c$,
at which point the dynamics becomes stable in the ``paramagnetic''
phase because the collective
decision $\sum_i s_i$ and therefore the market return have little
or no relationship with the external news. Hence, the term $\beta
r(t+1) G(t)$ takes random signs from one time step to the next,
leading to an effective random forcing added to the autoregressive
equation $K_i(t) = b_i + \alpha K_i(t-1)$.
We thus expect Gaussian distributions of returns
when $b_i/(1-\alpha)$ is smaller than $K_c$ and bimodal
distributions when $b_i/(1-\alpha) > K_c$ reflecting the slaving
of the global opinion to the sign of the news. Our
simulations, which have scanned 480 different models for
$b_{\max}$ from $0.1$ to $0.5$
with spacing $0.1$, $\sigma_{\max}$ from $0.005$ to $0.08$ with
spacing $0.005$, and $CV$ from $0.1$ to $1.1$ with spacing $0.2$,
confirm this prediction. Consider the distribution of returns
$r_{\tau}(t) = \ln[p(t)/p(t-\tau)]$ at
different time scales $\tau$.
For large idiosyncratic noise (large $CV$) and
not too large $b_{\max}$, the distribution of returns is Gaussian
for all time scales $\tau$. For smaller $CV$'s and larger
$b_{\max}$, we observe multimodal return distributions.
In the parameter space that we have explored and notwithstanding
our best attempts, we have not been able to find a set of
parameters leading to distributions of returns exhibiting a
monomodal shape with fat tails for small time scales, evolving
slowly towards Gaussian distributions at large time scales, as can
be observed in empirical data \cite{Mantegna-Stanley-2000}.
In addition, the correlation function of returns
($C_{\tau}(r,r)$) and of volatilities ($C_{\tau}(|r|,|r|)$) have
similar amplitudes and decay with the same characteristic time
scale as a function of time lag. This is very different from the
observed correlations of financial markets, with very short memory
for returns and long-memory for the volatility.

\begin{figure}[h]
\includegraphics[width=7cm]{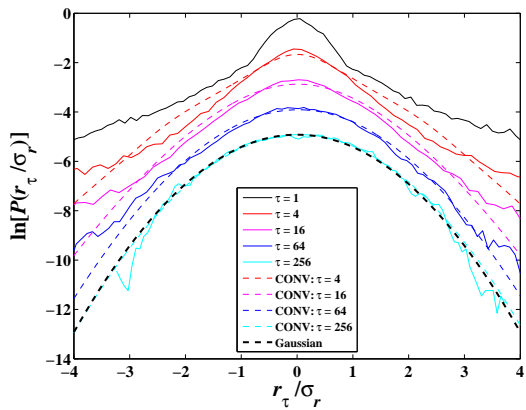}
\caption{(Color online) Probability distribution density (in
logarithmic scales) of log-returns at different time scales $\tau$
of a price time series obtained with $\lambda=40$, $\alpha=0.2$,
$b_{\max}=0.3$, $\sigma_{\max}=0.03$, and $CV=0.1$. The
log-returns $r_{\tau}$ are normalized by their corresponding
standard deviations $\sigma_{\tau}$. The pdf curves are translated
vertically for clarity. The thick dashed line is the Gaussian pdf.
The other dashed lines have been obtained by $\tau$-fold
convolutions of the pdf of the one-time step return $r_1(t)=\ln
p(t)/p(t-1)$.} \label{Fig1}
\end{figure}

\begin{figure}[h]
\includegraphics[width=7cm]{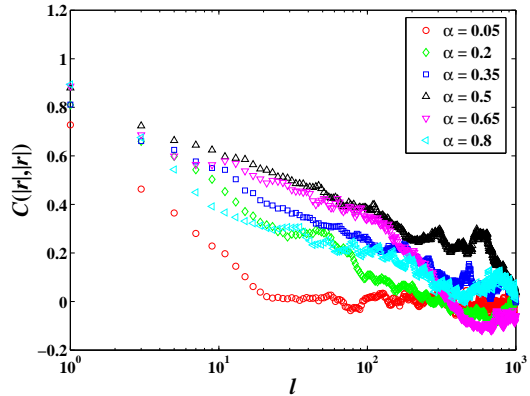}
\caption{(Color online) Impact of $\alpha$ on the auto-correlation
of the absolute values of the returns.} \label{Fig2}
\end{figure}

\begin{figure}[h]
\includegraphics[width=7cm]{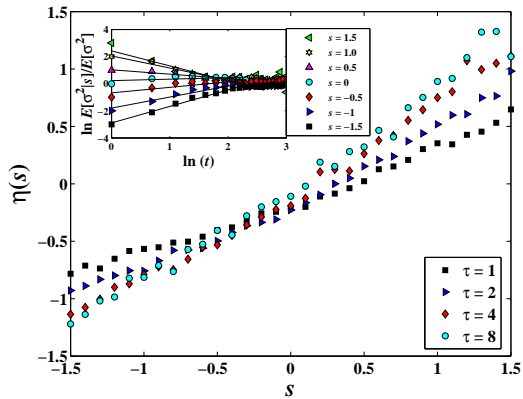}
\caption{(Color online) For a given relative log-amplitude $s$ of
a local burst of volatility occurring at some time $t_s$, we
translate and superimpose all time series starting at those times
of local bursts of the same amplitude $s$. Averaging over these time series of
volatility obtains the average conditional relaxation function of
the volatility $E[\sigma^2(t)|s] \sim t^{-\eta(s)}$ following a
local burst of volatility of amplitude $\propto e^s$. The inset
shows the average normalized conditional volatility
$E[\sigma^2|s]/E[\sigma^2]$ as a function of the time after the
local burst of volatility for different log-amplitudes $s$. The
figure shows the exponents $\eta(s)$ measured as the slopes of the
curves in the inset for $\tau=1, 2, 4$, and $8$. } \label{Fig3}
\end{figure}

For $\beta=1$, we obtained the following main stylized
facts of financial stock markets:
 (i) distributions of returns at different time scales $\tau$
(Fig.~\ref{Fig1}); (ii) correlation function of returns and of the
absolute value of the returns (Fig.~\ref{Fig2}); (iii) scaling of
the moments of increasing orders of the absolute values of the
returns (testing multifractality); (iv) the existence of a
hierarchy of exponents controlling the relaxation of the
volatility after an endogenous shock, another hallmark of
multifractality (Fig.~\ref{Fig3}); (v) the existence of bubbles
and crashes. We have explored 160 models with $b_{\max}$ from
$0.1$ to $0.5$ with spacing $0.1$, $\sigma_{\max}$ from $0.01$ to
$0.08$ with spacing $0.01$, and $CV$ from $0.1$ to $0.7$ with
spacing $0.2$. For each model, we generate time series of length
equal to $10^5$ time steps. We have found several parameter
combinations which lead to realistic stylized facts, for instance,
$(b_{\max}, \sigma_{\max}, CV)$ equal respectively to (0.3, 0.03,
0.1), (0.4, 0.04, 0.1), (0.4, 0.05, 0.1), (0.5, 0.06, 0.1), (0.1,
0.01, 0.3), (0.1, 0.02, 0.3), (0.2, 0.02, 0.3), (0.2, 0.03, 0.3),
(0.3, 0.04, 0.3), (0.5, 0.05, 0.3), (0.5, 0.07, 0.3), (0.3, 0.03,
0.5), and (0.5, 0.05, 0.5). The results presented here are for
($b_{\max}=0.3, \sigma_{\max}=0.03, CV=0.1$) which is typical.

Figure \ref{Fig1} shows the evolution of the pdf's of returns from
stretched exponential or power laws at short time scales that
cross over smoothly to a Gaussian law at the largest shown time
scale, in excellent agreement with empirical facts
\cite{Mantegna-Stanley-2000}. Note the difference between the
continuous and dashed lines for $\tau=4$, $16$, and $64$, which
expresses the existence of significant dependence in the time
series of returns. Such behavior is very similar to what is
observed in real data.

The temporal correlation of the log returns $r_1$ as a function of
the time lag $\ell$ exhibits a very short correlation time, of
duration smaller than one time step (not shown). In contrast, the
temporal correlation of the absolute value of log returns $r_1$
(``volatility''), taken as a proxy for the volatility, exhibits
long-range dependence up to approximately 1000 time steps. The
linear-log relationship suggested by the plots in Fig.~\ref{Fig2}
are predicted by the multifractal random walk (MRW) \cite{M_etal},
which provides an excellent model of many properties of financial
time series \cite{Luxmu}. The MRW depends only on three
parameters: the multifractal parameter $\lambda^2 \approx
0.02-0.04$, the integral time scale $T \approx 1$ year and the
standard deviation of returns. Comparing the dependence properties
of the returns and of the volatility suggests that one trading day
corresponds roughly to $5$ time steps of the model. This
correspondence translates into a integral time scale $T$ of about
$200$ days, which is compatible with empirical estimates for the
MRW \cite{MuzyQF}. The MRW also predicts (and this is
well-verified by empirical data) that the autocorrelation
functions of $|r_\tau(t)|$ for different $\tau$ should superimpose
for time lags larger than their respective $\tau$ \citep{M_etal}.
This prediction is also approximately observed in our model (not
shown).

Another important stylized facts is the multifractal structure of
the absolute values of log-returns \cite{Mandel97,MuzyQF}. We
verify the existence of a strong multifractality in our time
series (not shown) expressed by the scaling of the structure
function $M_q(\tau) \equiv \langle |r_{\tau}|^q \rangle \sim
\tau^{\xi_q}$, with exponents $\xi_q$ exhibiting a clear nonlinear
dependence as a function of the order $q$ of the structure
function. Rather than showing this standard looking multifractal
spectrum, we show in Fig.~\ref{Fig3} another striking signature of
multifractality discovered first in empirical data \cite{endovol}:
the MRW predicts a continuous spectrum of exponents $\eta(s)$ for
the relaxation of the volatility $E[\sigma^2(t)|s] \sim
t^{-\eta(s)}$  from a local peak as a function of its amplitude
$\propto e^s$ given by \be \eta(s)=\frac{2s}{3/2+\ln(T/\tau)}~.
\label{Eq:alpha:Ana} \ee

We have outlined an Ising model with imitation between agents,
their influence by external news and the impact of their private
information, which describes the ``digestion'' of  external news by the
collective behavior of the population of traders to create time
series of returns presenting long-range memory in the volatility
and multifractal properties, in agreement with empirical data.
This formulation has allowed us to test within the same model the
hypothesis that agents are rational versus irrational. The
empirical stylized facts of financial stock markets have been
found only when agents misinterpret, or mis-attribute the source
of the prediction of returns or are over-confident ($\beta >0$). We can
interpret our results by saying that, conditioned on their role of
reflecting the stock market, the news serve as the substrate for
fostering social interactions and reinforcing herding.
Technically, the stylized facts in this regime result from the
fact that the model operates around the critical point of the
corresponding Ising model, with coupling coefficients which are
time-dependent and endowed with a memory of past realizations. The
critical point of the Ising model is associated with a critical
value $K_c$ for the average coupling coefficient. Close to this
value, agents organize spontaneously within clusters of similar
opinions, which become very susceptible to small external
influences, such as a change of news. This may explain the
occurrence of crashes as argued previously\cite{Sornette-2003}.
Intuitively, the critical slowing down well-known to characterize
the proximity to the critical Ising point can explain the
long-term memory of the volatility while the almost absence of
correlation of the returns themselves is ensured by the impact of
the news and the random idiosyncratic decisions.

As a bonus, we have discovered that this simple model exhibits a
rich multifractal structure, diagnosed not only by the standard
convexity of the exponents of the structure functions but also by
a continuous spectrum of power law response functions to endogenous
shocks \cite{endovol,endoamazon}. To
our knowledge, this is the first nonlinear model in which such
clear distinction is documented quantitatively, based on a
bottom-up self-organization. In contrast, the multifractal random
walk which has provided the theoretical predictions used here is a
descriptive phenomenological model.

\end{document}